**A comparison of the resistivity behavior of $MgB_2$, $AlB_2$ and $AgB_2$ systems**

R. Lal[1], V.P.S. Awana[1,$], K. P. Singh[1], H. Kishan[1] and A.V. Narlikar[2]

[1]National Physical Laboratory K.S. Krishnan Marg, New Delhi 110012, India

[2]UGC-DAE Consortium for Scientific Research, University Campus, Khandwa Road, Indore-452017, India

Measurements have been performed of the resistivity of the samples of $MgB_2$, $AlB_2$ and $AgB_2$. The samples show presence of impurities. Analyzing the data in terms of the impurity scattering, electron-phonon scattering, and weak localization it has been found that the $AlB_2$ ($AgB_2$) sample involves maximum (minimum) effect of the impurity, electron-phonon interaction and weak localization.

Key words: Di-borides, Resistivity measurements, Conduction mechanism

[$] Corresponding Author; e-mail: awana@mail.nplindia.ernet.in

**INTRODUCTION**

$MgB_2$ is not a new compound, it has been known since the early 1950's. Only in 2001 it was discovered to be a superconductor at a remarkably high critical temperature of about 39K [1]. This discovery stimulated global interest seeking higher $T_c$ and uncovering the basic physics [2]. $MgB_2$ attracts a lot of attention from both condensed matter experimentalists and theorists [1, 2, 3]. Its high $T_c$ comes from the exceptionally high vibrational energies in the graphite-like boron planes and thus $MgB_2$ appears to obey conventional models of superconductivity. This relatively simple view (as compared to HTS) opens up a wide range of practical opportunities [2]. Based on various physical property measurements, important critical parameters of the compound viz., critical superconducting temperature ($T_c$), coherence length ($\xi$), penetration depth ($\lambda$), critical current ($J_c$) and lower/upper critical fields $H_{c1}/H_{c2}$ have already been determined and reviewed [4, 5]. Compared to HTS, $MgB_2$ possesses simpler structure, lower anisotropy and larger coherence length. Most interestingly $MgB_2$ has nearly transparent grain boundaries [6], which permit excellent current transport. Higher quality grain boundaries and better superconducting critical parameters provide $MgB_2$ an edge over widely studied HTSC cuprates [7].

MgB2 has a simple hexagonal $AlB_2$-type structure (space group P6/mmm), which is common among borides. It contains graphite-type boron layers, which are separated by hexagonal close-packed layers of magnesium. The magnesium atoms are located at the center of hexagons formed by boron and donate their electrons to the boron planes [1, 2]. Similar to graphite, $MgB_2$ exhibits anisotropy in the B-B length: the distance between the boron planes is significantly larger than that in-plane B-B distance [1, 2]. Among several di-borides $AgB_2$ and $AuB_2$ were thought to be good candidates for higher $T_c$ [8]. There exists only one report of the observation of superconductivity of up to 6.7 K in a laser ablated $AgB_2$ thin film [9]. This value of $T_c$ is remarkably lower than theoretically predicted value of 59 K [8]. In fact the mechanical behavior of AgB2 was studied as a function of pressure in comparison to $MgB_2$ superconductor [10]. AgB2 was found to be a much more tightly packed incompressible material compared to $MgB_2$.  $AlB_2$ is the parent compound of the di-borides family, and is not predicted to be superconducting. In this note, we report the synthesis of phase pure $MgB_2$, $AlB_2$ and $AgB_2$ compounds and analyze critically their resistivity behavior down to 12 K in terms of impurity scattering, electron-phonon scattering, and weak localization. It is found that the $AlB_2$ ($AgB_2$) sample involves maximum (minimum) effect of the impurity, electron-phonon interaction and weak localization.

**EXPERIMENTAL DETAILS**

Various di-borides viz $MgB_2$, $AlB_2$ and $AgB_2$ compounds were synthesized using high quality Mg, Al, Ag and B powder, by mixing them in stoichiometeric ratio. The mixed and ground powder, are further palletized. The pellets are than put in closed end soft iron (SS) tubes. The pellets containing SS tubes were than sealed inside a quartz tube at high vacuum of $10^{-5}$ Torr, please see Fig.1. The encapsulated raw pellets are than heated at 750 $^0$C with a hold time of 3 hours and finally quenched in liquid nitrogen ($LN_2$).  X-ray diffraction (XRD) patterns were obtained at room temperature using $CuK_\alpha$ radiation. Resistivity measurements were made in the temperature range of 12 to 300 K using a four-point-probe technique on a Close Cycle refrigerator (CCR).

**RESULTS AND DISCUSSION**

Various di-borides viz. $MgB_2$, $AlB_2$ and $AgB_2$ compounds crystallize in Hexagonal ($P_{6/mmm}$) structure without any noticeable impurity (plots not shown). The lattice parameters are a = 3.08 Å and c = 3.51 Å for $MgB_2$, a = 3.01 Å and c = 3.24 Å for $AlB_2$ and a = 2.88 Å and c = 3.53 Å for $AgB_2$.   In Figs. 2-4 we show the resistivity of $MgB_2$, $AlB_2$ and $AgB_2$ up to room temperature. While there is superconducting phase transition in $MgB_2$ at $T_c$= 38 K, the low temperature behavior of the resistivity of $AlB_2$ and $AgB_2$ show an effect of localization. Since the main interaction in the studied diboride is the electron- phonon interaction [11-13], and more over the metallic nature of resistivity for large $T$ suggests a weak localization, resulting in

$$\rho(T) = \rho_o + A \frac{T^5}{\theta^4_D} \int\limits_{0}^{\theta_D/T} \frac{z^5}{(e^z-1)(1-e^{-z})} dz - B \ln T \qquad (1)$$

Here the first term gives the temperature independent contribution to resistivity due to the impurities present in the system. The second term arises due to the contribution of the electron-phonon interaction within the Bloch-Gruneisen theory [14]. In this term $A$ is a constant independent of both temperatures $T$ and Debye temperature $\theta_D$. The third term is the contribution due to the weak localization [15]. Here $B$ is a constant. While the constant $A$ provides the contribution of the electron-phonon interaction, the constant $B$ provides the contribution of the weak localization.

We have fitted the resistivity of $MgB_2$, $AlB_2$ and $AgB_2$ with Eq. (1). The value of the parameters $\rho_o$, $A$, $\theta_D$ and $B$ are given in table I. We emphasize that the set of values $\rho_o$, $A$, $\theta_D$ and $B$ for a given sample is unique. This is because $\rho_o$, $B$ and $(A, \theta_D)$ correspond to qualitatively much different functional dependence of $\rho(T)$, and $(A, \theta_D)$ govern the $low - T$ curvature and $high - T$ linear variation of $\rho(T)$. So, $\rho_o$, $B$ and $(A, \theta_D)$ will be unique. As far as the separate uniqueness of $A$ and $\theta_D$ is concerned, it may be noted that while the $low - T$ curvature of $\rho(T)$ [excluding the upturn] depends upon $A/\theta_D{}^4$, the $high - T$ linear part of $\rho(T)$ depends on A only. This means the value of $\theta_D$ will also become unique. In this way the set of values $\rho_o$, $A$, $\theta_D$ and $B$, given in table I, is a unique set, and no other set of values of these parameters can provide an equally well fit of the experimental data with Eq. (1).

From the values of $\rho_o$ and $A$ it appears that the effect of impurities and electron-phonon interaction is maximum in the AlB2 sample, while it is least in the $AgB_2$ sample. As far as the contribution of the weak localization is concerned, values of $B$ show no weak localization in $MgB_2$. In comparison to $AgB_2$ the effect of weak localization is about 20 times larger in $AlB_2$. This is consistent with the values of $\rho_o$ for these two samples, because for $AlB_2$ $\rho_o$ is about 22 times larger than $\rho_o$ of $AgB_2$. The Debye temperature is found to decrease in the order $\theta_D(MgB_2) > \theta_D(AlB_2) > \theta_D(AgB_2)$. Since one of the factors on which $\theta_D$ depends is the mass of the constituent atoms, and since mass (Mg) < mass (Al) < mass (Ag), it may be argued that the atoms of Mg, Al and Ag make essential contribution in determining the Debye temperature.

In conclusion, we have synthesized sample of $MgB_2$, $AlB_2$ and $AgB_2$, and have measured their resistivity up to room temperature. The combined effect of the impurity scattering, electron-phonon interaction (Bloch-Gruneisen theory) and weak localization provides a reasonable explanation of the resistivity data with various parameter value given in table I.

**Table I:** Values of the parameters $\rho_o$, $A$, $\theta_D$ and $B$ which appear in Eq.(1) for the system $MgB_2$, $AlB_2$ and $AgB_2$.

| System | $\rho_o$ ($\mu\Omega$cm) | $A$ ($\mu\Omega$cm) | $\theta_D$ (K) | $B$ ($\mu\Omega$cm) |
|---|---|---|---|---|
| $MgB_2$ | 10.3 | 0.4 | 700 | 0.0 |
| $AlB_2$ | 26.8 | 0.8 | 670 | 0.02 |
| $AgB_2$ | 1.18 | 0.082 | 480 | 0.001 |

**FIGURE CAPTIONS**

Fig.1. Photograph of SS tube encapsulated raw $MgB_2$ compound at $10^{-5}$ torr.

Fig.2. $\rho$ (T) of $MgB_2$ compound

Fig.3 $\rho$ (T) of $AlB_2$ compound

Fig.4 $\rho$ (T) of $AgB_2$ compound

Fig. 1

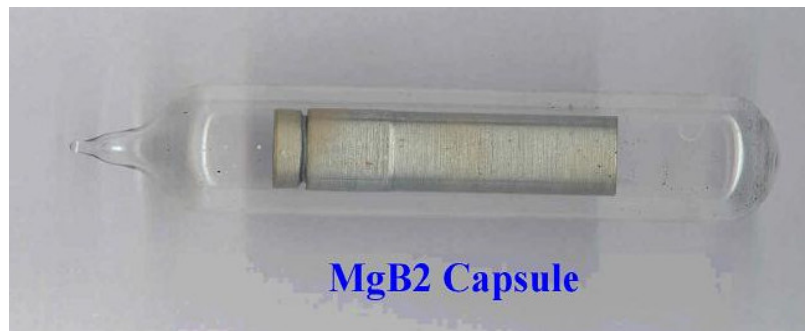

**MgB2 Capsule**

Fig. 2

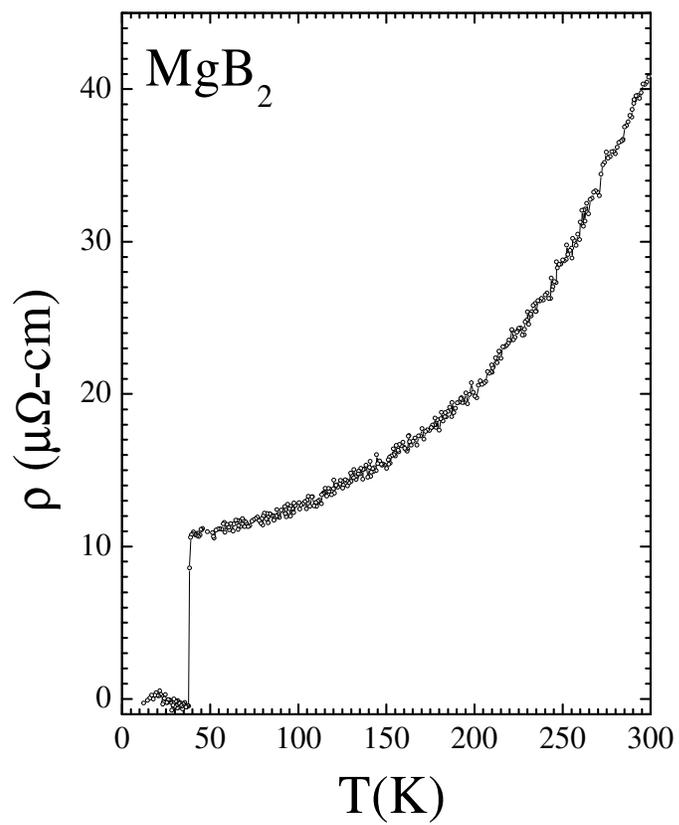

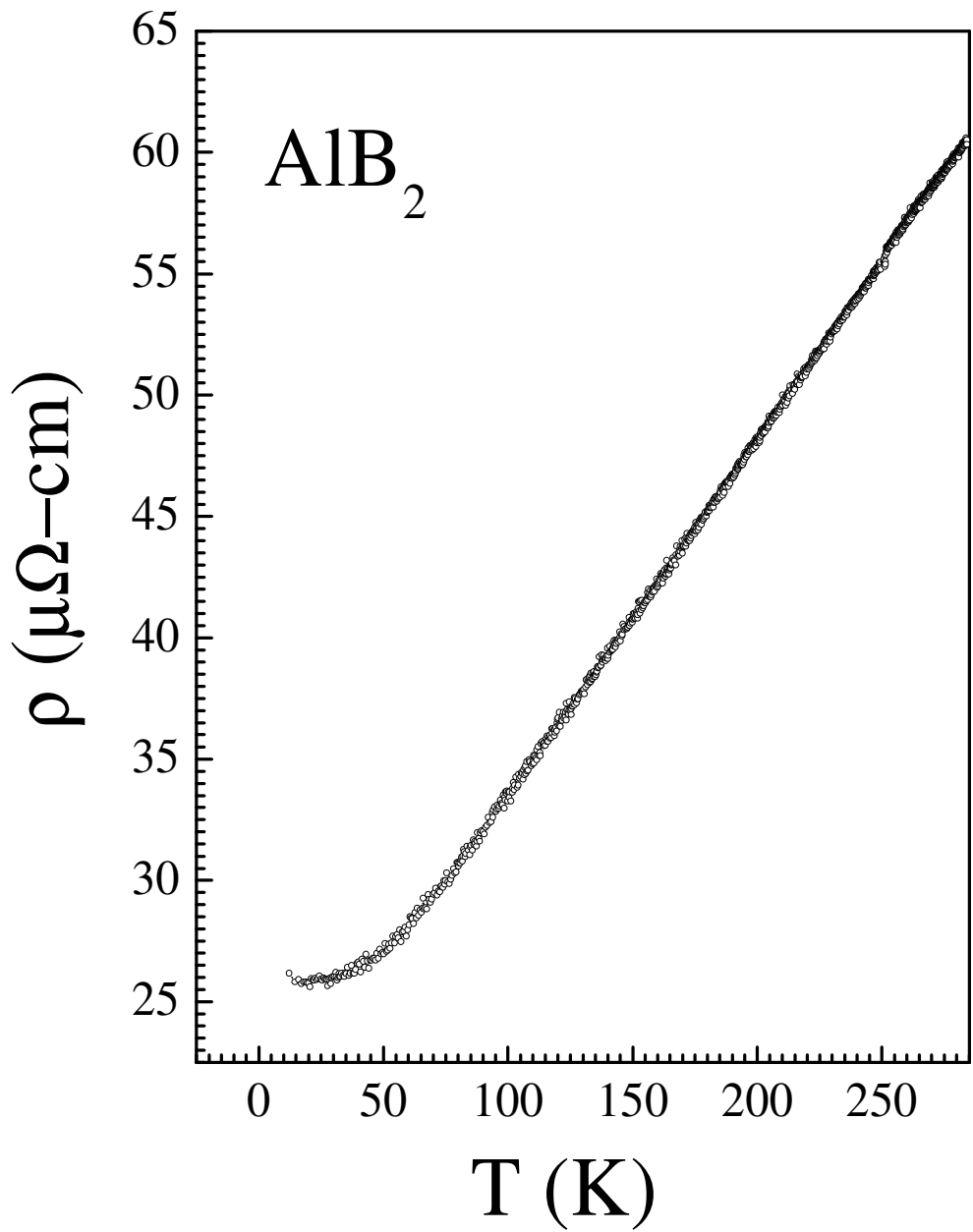

AlB$_2$

Fig.4

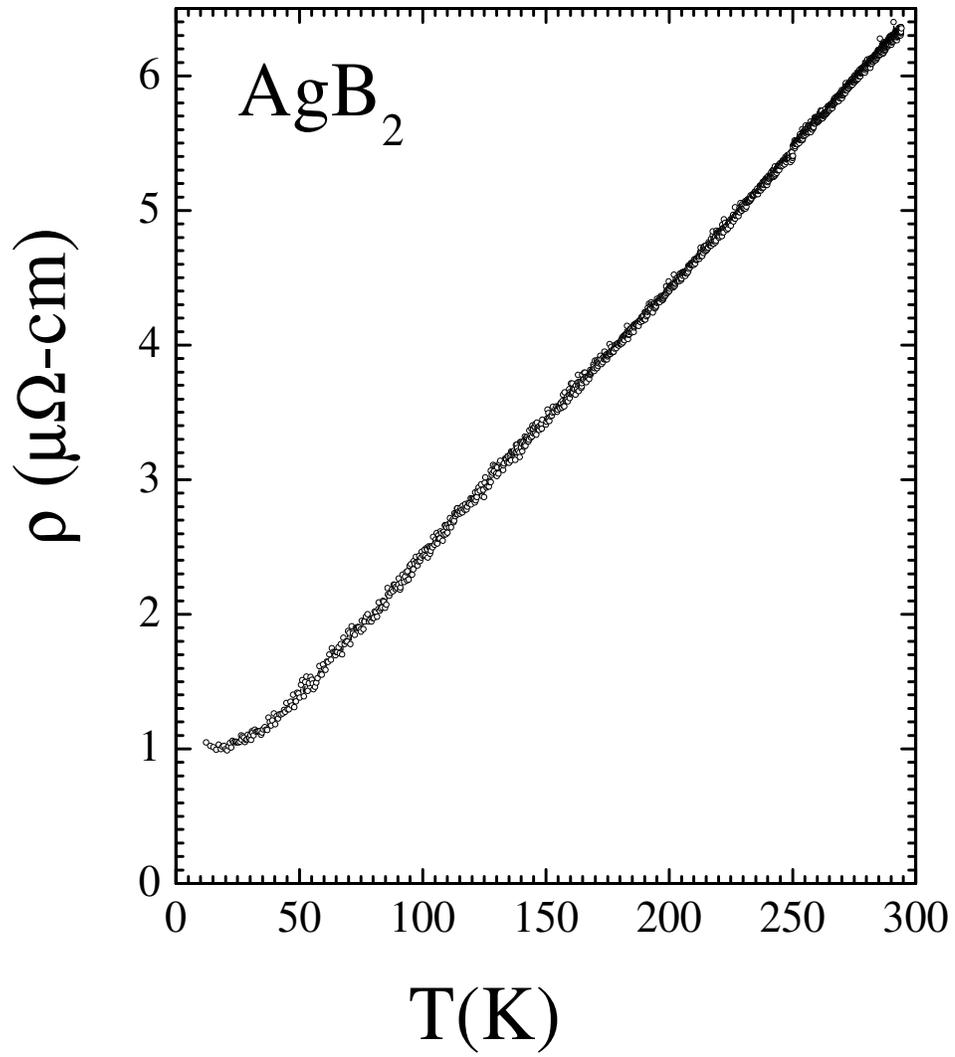